**Marriage Discourse on Chinese Social Media: An LLM-assisted Analysis**

Frank Tian-Fang Ye[1] and Xiaozi Gao[2]

[1]Division of Social Sciences, The HKU SPACE Community College, Hong Kong SAR, P.R.C.
https://orcid.org/0000-0001-7248-3062
Email: frank.ye@hkuspace.hku.hk

[2]Corresponding Author, Department of Early Childhood Education, The Education University of Hong Kong, Hong Kong SAR, P.R.C.
https://orcid.org/0000-0003-1452-7957
Email: gaox@eduhk.hk

## Abstract

China's marriage registrations have declined substantially, dropping from 13.47 million couples in 2013 to 6.1 million in 2024. This study examined sentiment and moral elements underlying 219,358 marriage-related posts from Weibo and Xiaohongshu using large language model (LLM)-assisted content analysis. Drawing on Shweder's Big Three moral ethics framework, posts were coded for sentiment (positive, negative, neutral) and moral elements (autonomy, community, divinity). Results revealed platform differences: Weibo leaned toward positive sentiment, while Xiaohongshu was predominantly neutral. Most posts lacked explicit moral framing. However, when moral elements were invoked, significant associations with sentiment emerged. Posts invoking autonomy and community were predominantly negative, whereas divinity-framed posts tended toward positive sentiment. These findings suggest that concerns about both personal autonomy constraints and communal obligations contribute to negative marriage attitudes in contemporary China, offering insights for culturally informed policies addressing marriage decline.

**Marriage Discourse on Chinese Social Media: An LLM-assisted Analysis**

## Introduction

Trends toward delayed and declining marriage have been pronounced in East Asian societies (Raymo et al., 2015). In China, marriage registrations have declined from a peak of 13.47 million couples in 2013 to approximately 6.1 million in 2024, representing a 20.5% year-on-year decline from 2023 and the lowest figure since 1980 despite a larger total population (National Bureau of Statistics, 2025). The crude marriage rate has dropped to 4.3 per 1,000 population, comparable to the demographic patterns observed in Japan and South Korea (Yi, 2025). This decline has implications for fertility rates, particularly in China where childbearing remains closely tied to marriage (Yu & Xie, 2022). Understanding how contemporary Chinese society perceives marriage is therefore crucial for addressing these demographic challenges.

Existing research on marriage attitudes, however, faces three main limitations. First, while studies have examined moral reasoning in relation to specific marriage-related issues, such as attitudes toward same-sex marriage (Koleva et al., 2012) and judgments about relationship norm violations (Selterman & Koleva, 2015), they focused primarily on specific issues (marriage equality, non-monogamy) rather than sentiment toward marriage as an institution broadly. Understanding the moral foundation underlying general attitudes toward marriage is essential because sentiment alone reveals whether people view marriage positively or negatively, while moral reasoning reveals why these sentiments are held and what ethical principles justify them. This distinction matters because the same negative sentiment may arise from fundamentally different moral elements (e.g., autonomy-based concerns about constraint vs. community-based concerns about oppressive obligations), and these different moral elements require different policy or educational approaches.

Second, most existing research on marriage attitudes relies on survey methodologies (Yeung & Hu, 2016), which, despite their value, have inherent limitations. Survey responses can be influenced by question wording and framing effects and affected by social desirability bias wherein respondents provide socially acceptable rather than authentic answers (Tourangeau & Yan, 2007). Additionally, surveys capture how people respond when directly asked but not how they spontaneously express and morally justify their views about marriage in naturalistic contexts.

Third, prior research concentrates on Western or predominantly Christian contexts (Koleva et al., 2012), leaving the moral elements of marriage attitudes in rapidly changing non-Western societies largely unexplored. Given that East Asian societies are experiencing significant marriage rate declines globally while simultaneously balancing cultural tensions between traditional values emphasizing collective obligations and modern individualistic orientations, examining this issue in the Chinese context is particularly necessary.

Social media platforms offer a promising avenue for addressing these limitations by providing access to spontaneous, self-generated public discourse. When individuals post about marriage on social media, they do so in their own words, without researcher prompting, and in contexts they choose (e.g., celebratory announcements, venting frustrations, seeking advice). Importantly, such discourse reveals not only what people feel about marriage (emotional sentiment) but also how they morally reason about it.

To address these gaps, we conducted a large-scale content analysis of Chinese social media discourse about marriage. We collected approximately 220,000 posts from two major platforms: Sina Weibo and Xiaohongshu. Drawing on Shweder et al. 's (1997) Big Three morality framework, which identifies three universal moral elements (autonomy, community, and divinity), we coded posts for both moral element and emotional sentiment. To handle this data volume, we employed large language models (LLMs) to assist with systematic coding while maintaining rigorous validation procedures.

We addressed four research questions (RQ): First, what sentiments (positive, negative, neutral) are expressed in marriage-related social media posts? Second, what moral elements (autonomy, community, divinity) are invoked in these posts? Third, how do expressions of moral elements relate to sentiment valence? Fourth, what topics co-occur with marriage discourse in social media discussions? Our findings offer insights for developing culturally informed policies and communication strategies responsive to the concerns driving marriage decline in East Asian contexts.

**The Emotional and Moral Elements of Marriage Discourse**

Contemporary discourse about marriage on social media reveals diverse attitudes, ranging from celebratory engagement announcements and praise for marital partnership to critiques of traditional marriage expectations and expressions of skepticism (Nabilah et al., 2024). Recent empirical studies document this complexity. Research examining Chinese social media found that discussions about women's marriage and fertility exhibited predominantly negative sentiment, driven by concerns about societal pressures, gender inequality, and work-family conflicts (He et al., 2024). Similarly, analysis of marriage-related hashtags on TikTok revealed that most comments expressed negative views, reflecting widespread anxieties about marriage's implications for personal autonomy and well-being (Nabilah et al., 2024). This negativity tendency across platforms suggests that negative and ambivalent perspectives may be particularly salient in shaping contemporary marriage discourse.

These attitudes toward marriage are not merely personal preferences or emotional reactions; they are grounded in moral reasoning about what is right, good, and proper in human relationships and social life. Research demonstrates that moral concerns play a crucial role influencing how people evaluate marriage-related issues. For instance, studies have found that moral foundations, particularly purity/sanctity, strongly predict attitudes toward same-sex marriage (Koleva et al., 2012; Ochoa et al., 2016), while different moral concerns influence judgments about relationship behaviors such as infidelity, non-monogamy, and commitment (Selterman & Koleva, 2015). These findings suggest that when people evaluate marriage, they draw on deeper moral concerns about individual rights, social obligations, and sacred values. Understanding contemporary marriage attitudes therefore requires examining not only whether people view marriage positively or negatively, but also the moral logics that underlie and justify these evaluative stances.

Shweder and colleagues' (Shweder et al., 1997) Big Three ethics framework provides a straightforward approach for understanding the moral elements of marriage attitudes. This framework identifies three distinct moral orientations: the ethic of autonomy, which centers on individual rights, personal freedom, harm prevention, and fairness; the ethic of community, which emphasizes social roles, duties, hierarchy, and collective welfare; and the ethic of divinity, which focuses on sanctity, purity, and sacred values (Jensen, 2011). Marriage represents an appropriate domain for examining all three elements because it simultaneously implicates individual choice and well-being (autonomy), family obligations and social roles (community), and sacred traditions and values (divinity).

Empirical research demonstrates how these distinct moral elements shape marriage attitudes and behaviors in different ways. Autonomy-oriented values predict higher justification for divorce and emphasis on personal fulfillment, though excessive focus on individual autonomy can increase marital conflict (Giudici et al., 2011; Mentser & Sagiv, 2025). Community-oriented values frame marriage as a socially embedded institution centered on family duty rather than personal satisfaction, creating relationship patterns and well-being outcomes differ from individualistic approaches (Bejanyan et al., 2015; Ten Kate, 2015). Divinity-oriented perspectives frame marriage as sacred, with sanctification buffering against relationship distress, though rigid purity ideologies can negatively affect marital satisfaction (DeMaris et al., 2010; Sawatsky et al., 2024). These findings indicate that the same evaluative stance to marriage, be it positive or negative, can arise from different moral concerns.

**Marriage in Contemporary China**

China's marriage decline unfolds within a unique cultural context characterized by tensions between traditional collectivist values and emerging individualistic orientations. Historically, Chinese marriage practices have been embedded in Confucian frameworks that emphasize filial piety, hierarchical authority, and the continuation of family lineages (Raymo et al., 2015). Within this tradition, marriage was primarily understood through the ethic of community, as a means of fulfilling family duties and maintaining social order rather than expressing individual choice or pursuing romantic fulfillment.

However, rapid socioeconomic transformation since China's economic reforms has introduced significant value shifts. Research documents rising individualism among Chinese youth, who increasingly prioritize personal goals, emotional fulfillment, and self-actualization over traditional family obligations (Yeung & Hu, 2016). This shift reflects a growing emphasis on the ethic of autonomy, as young adults seek marriages based on personal compatibility and mutual affection rather than family arrangement or social duty. Increasingly, marriage is framed as a matter of individual choice and romantic love, with young people expecting emotional intimacy and egalitarian partnerships rather than hierarchical relationships defined by duty.

Yet these individualistic values coexist uneasily with persistent traditional expectations, creating what Yeung and Hu (2016) term a "paradox" in which marriage ideals and behaviors increasingly diverge. Chinese women particularly face intense pressure regarding marriage timing and childbearing, with media discourse stigmatizing "leftover women" (sheng nu) who remain unmarried past their mid-twenties (To, 2013). Paradoxically, while individualization has expanded women's educational and economic opportunities, it has also reinforced rather than eliminated traditional gender norms within marriage (Xie, 2021). Young Chinese adults thus find themselves caught between competing demands: they desire autonomy-based marriages while simultaneously facing family pressure to fulfill community-based duties through timely marriage and childbearing. Examining how Chinese social media users invoke different moral concerns when discussing marriage therefore provides crucial insight into these cultural tensions and their contributions to China's demographic challenges.

**Marriage Discourse on Social Media Platforms**

Over the past decade, social media platforms have emerged as valuable data sources for studying public opinion and mechanisms shaping attitudes across diverse topics. For example, studies use Twitter (X) data to track political sentiment during elections (Balasubramanian et al., 2024), health crisis perceptions (Thakur et al., 2023), and policy discourse such as universal basic income debates on Reddit (Kim et al., 2025). Chinese social media platforms offer parallel insights: Sina Weibo, for instance, has been used to analyze COVID-19 vaccine attitudes (Gao et al., 2022), public psychology during pandemic waves (Pan et al., 2021), and perceptions of online privacy following privacy breach events (Lee et al., 2022). In the domain of marriage and family, researchers have similarly employed natural language processing and sentiment analysis to capture public attitudes. These studies document real-time reactions to marriage-related legal decisions, track linguistic shifts following engagement announcements, and reveal variations in marriage expectations (He et al., 2024; Liu, 2024; Nabilah et al., 2024). Such approaches enable researchers to access marriage-related attitudes that individuals may not readily express in traditional surveys or face-to-face interviews.

Beyond documenting sentiment, research has shown that language reflecting moral foundations appears prominently in social media discussions of diverse topics, including vaccination (Borghouts et al., 2023), mental health (Mittal & De Choudhury, 2023), immigration (Grover et al., 2019), and climate change (Song et al., 2025). This suggests that moral reasoning frameworks like the Big Three ethics provide relevant lenses for analyzing marriage discourse. Yet research has rarely examined marriage attitudes through this moral framework on social media.

**The present study**

The present study addresses this gap by analyzing marriage discourse on two major Chinese platforms: Weibo and Xiaohongshu. These platforms represent important spaces for public discourse in China, offering complementary windows into how individuals reason about marriage. Weibo, with its broader public reach, facilitates large-scale debates on social issues, while Xiaohongshu, with its emphasis on personal lifestyle content, enables more intimate personal reflection and experience-sharing. Together, these platforms capture both public argumentation and private contemplation about marriage. Guided by the Big Three framework, we analyzed large-scale social media discourse to examine how moral elements and sentiment are expressed in naturalistic contexts. We addressed four RQs.

**RQ1**: What sentiments are expressed in marriage-related social media posts? Based on recent findings that Chinese social media marriage discourse is predominantly negative and reflects anxieties about marriage's implications for autonomy and wellbeing (He et al., 2024; Nabilah et al., 2024), we hypothesized that negative sentiment would predominate in marriage-related posts, particularly on Weibo where public debate about social issues is more prevalent.

**RQ2**: What moral elements are invoked in marriage-related social media posts? Given China's ongoing individualization and the increasing emphasis on personal fulfillment among younger generations (Yeung & Hu, 2016), we hypothesized that autonomy would be the most frequently invoked moral element in marriage discourse.

**RQ3**: How do expressions of moral elements relate to sentiment in marriage-related posts? Research linking autonomy concerns to marital dissatisfaction (Giudici et al., 2011; Kluwer et al., 2020) and divorce acceptance (Mentser & Sagiv, 2025) suggests that autonomy-based reasoning often accompanies critical stances toward marriage. In the Chinese context, where marriage traditionally involved sacrifices of individual freedom (particularly for women), we hypothesized that posts invoking autonomy elements would express predominantly negative sentiment.

The relationship between community elements and sentiment appeared more complex. Research shows that community values can provide protective effects in collectivistic contexts (Parry, 2016) but also create pressures that attenuate wellbeing benefits (Bejanyan et al., 2015; Ten Kate, 2015). In Chinese marriage discourse, community might accompany both positive sentiment (emphasizing family harmony, intergenerational bonds, and social stability) and negative sentiment (highlighting oppressive familial expectations, arranged marriage pressures, and loss of individual agency). We therefore hypothesized that community would be associated with more balanced or mixed sentiment patterns compared to autonomy.

Although divinity might be less prevalent, research on sanctification suggests that when marriage is viewed through sacred or transcendent lenses, it provides buffering effects against stressors (DeMaris et al., 2010). Posts invoking divinity might emphasize marriage's spiritual meaning, or sacred obligations in ways that frame marriage positively. We therefore hypothesized that posts expressing divinity would tend toward positive sentiment.

**RQ4**: What topics co-occur with marriage discourse in social media discussions? This exploratory research question aimed to understand what concepts and themes appeared alongside marriage-related terms in naturalistic discourse. Given its exploratory nature, we did not form specific hypotheses.

# Method

## LLM-Assisted Content Analysis

Analyzing the moral elements of marriage discourse at scale requires methodological approaches capable of processing large textual datasets while maintaining interpretive rigor. The emergence of large language models (LLMs) has introduced new possibilities for qualitative social science research, particularly in analyzing large-scale social media data. Recent studies demonstrate that LLMs can achieve human-equivalent performance on complex coding tasks, with GPT-4 reaching substantial to near-perfect intercoder reliability (Cohen's $\kappa \geq 0.79$) on socio-historical coding tasks (Dunivin, 2024). This methodological advancement enables researchers to scale qualitative analysis from hundreds to millions of texts while maintaining interpretive rigor through a hybrid workflow model.

Chinese-language LLMs have developed rapidly, with models such as Alibaba's Qwen series and Baidu's ERNIE demonstrating strong performance on Chinese natural language understanding benchmarks (Baidu-ERNIE-Team, 2025; Qwen Team, 2025). These models offer advantages for analyzing Chinese social media platforms, having been trained extensively on Chinese linguistic patterns and cultural contexts.

**Analytic Strategy**

Our analytical strategy proceeded in three stages. First, we filtered out irrelevant content such as advertisements, fictional stories, and official announcements. Second, we analyzed the sentiment and moral elements (based on the Big Three framework) expressed in marriage-related discussions. Third, we conducted topic modeling using word embeddings to identify topics and themes associated with marriage in the datasets.

To implement this strategy, we selected two open-source LLM models for automated content classification and coding: gpt-oss-20B (OpenAI, 2025) and Qwen3-32B. These models were chosen based on three considerations. First, both offer open-weight architecture with permissive licensing (Apache 2.0), enabling complete local deployment through Ollama. This addresses data privacy concerns and ensures reproducibility through fixed model versions without dependence on commercial API services (Bi et al., 2025). Second, both employ a Mixture-of-Experts (MoE) architecture and support advanced features critical for qualitative coding, including chain-of-thought reasoning, function calling, and structured output generation. These capabilities enable transparent and auditable classification decisions. Third, the 20-32 billion parameter range balances computational efficiency with accuracy for Chinese text analysis.

## Data Collection and Sampling

Data were collected from two major Chinese social media platforms, namely Weibo and Xiaohongshu using a custom web scraping program developed in Python with the Scrapy framework (Lei, 2020/2025; *Scrapy*, 2010/2025). Platform search functions were queried using six Chinese keywords representing marriage-related terminology: 婚姻 (marriage), 嫁娶 (marry/wedding), 成亲 (getting married), 成婚 (becoming married), 成家 (establishing a family), and 结婚 (to marry). These search terms captured varying registers and contexts in Chinese language usage, targeting publicly available posts related to marriage discourse in contemporary Chinese society.

### *Weibo Data Collection*

Over a five-week data collection period, we gathered 181,150 Weibo posts published between July 2024 and June 2025. Posts were fairly evenly distributed across months, ranging from 11,042 to 19,863 per month (M = 15,096). We first removed 76 posts (0.04%) that contained only hashtags without substantive content.

Preliminary examination revealed substantial advertisements and promotional content in the dataset, necessitating systematic identification and removal of non-genuine marriage discussions. To address this challenge, we implemented an LLM-assisted classification system using the gpt-oss-20B and Qwen3-32B models described in the Analytical Strategy section. The classification workflow was deployed through Dify (Dify, 2023/2025), a workflow automation platform with local model deployment via Ollama (Ollama, 2023/2025). Both models were configured with reasoning mode enabled, allowing them to generate intermediate reasoning steps before producing final classifications.

The classification workflow consisted of a three-node pipeline: User Input (social media content) → LLM Classification Node → Output (classification result). We employed a few-shot prompting approach with five hand-coded examples representing distinct content types. The task used a binary coding scheme in which posts were classified as YES (genuine marriage-related content) or NO (excluded content including advertisements, creative fiction, or official announcements). The system prompt directed the models to: (1) carefully read the complete post content, (2) identify key indicators for each category, (3) output only the classification label (YES or NO), and (4) default to YES when uncertain. Key indicators for exclusion included product descriptions, pricing, and promotional language for advertisements; narrative structure, character development, and chapter numbering for fiction; and formal notices, policy updates, and administrative tone for official announcements. Genuine content was characterized by personal experiences, opinions, and information sharing about marriage.

**Validation of LLM Classification**. To validate the LLM classification approach, we conducted an inter-rater reliability analysis using a randomly selected sample of 200 Weibo posts. Human coders served as the ground truth, with classifications compared against the two LLM models. The gpt-oss-20B model demonstrated high agreement with human raters (Cohen's $\kappa = 0.867$, $p < .001$), achieving 94.0% overall agreement. The confusion matrix (Figure 1) indicated high sensitivity (92.6%) and specificity (96.9%). In contrast, the Qwen3-32B model showed substantial but lower agreement (86.0% raw agreement, Cohen's $\kappa = 0.705$, $p < .001$), with notably reduced sensitivity (81.6%) due to a conservative classification bias that resulted in higher false negative rates.

Both models exhibited conservative classification tendencies, with false negative to false positive ratios of 5.0:1 (gpt-oss-20B) and 8.3:1 (Qwen3-32B). However, the gpt-oss-20B model's superior performance metrics, including precision of 0.984 for YES classifications and balanced F1-scores exceeding 0.91 for both classes, indicated its suitability for automated content classification. Based on these reliability findings, we employed the gpt-oss-20B model for automated classification of the full Weibo dataset, yielding a final sample of 71,988 posts.

**Figure 1**

*Confusion Matrices for Inter-rater Reliability Between Human Coders and LLM Models (gpt-oss-20B and Qwen3-32B) on Content Classification*

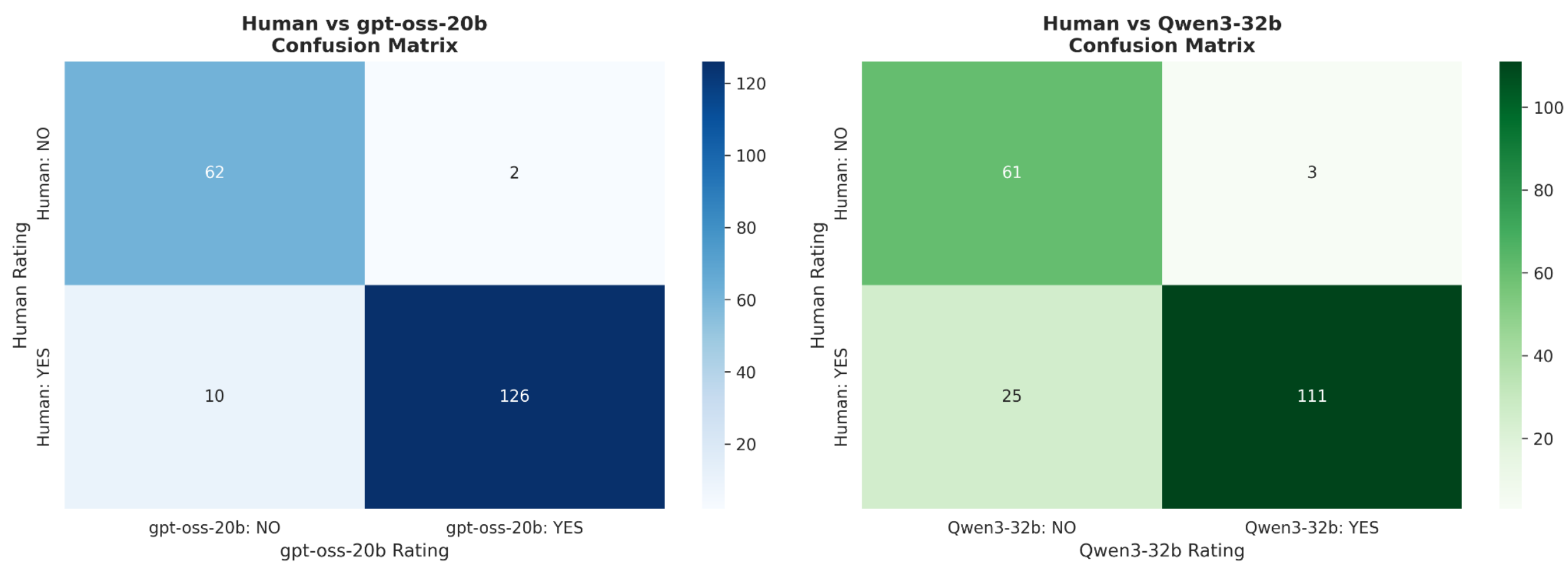

### *Xiaohongshu Data Collection*

Xiaohongshu data collection yielded an initial dataset of 37,017 posts. Due to restrictions imposed by the platform's search function, data were collected on a daily basis over a one-month period to ensure adequate sample coverage. Preliminary quality control revealed that 5,893 posts (15.92%) lacked substantive content and were removed. Deduplication procedures using post titles as the primary matching criterion identified 11,184 duplicate posts (35.93% of the cleaned dataset). Following these procedures, the final Xiaohongshu sample consisted of 19,940 unique posts.

### *Final Dataset*

All data processing procedures were conducted using custom Python scripts. Following all quality control procedures, the final analytical dataset comprised 71,988 Weibo posts and 19,940 Xiaohongshu posts (N = 91,928 total posts).

## Data analysis

### *Sentiment and Moral Element Analysis*

Sentiment and moral element analysis were conducted using the same LLM-assisted approach described in the Analytical Strategy section. For sentiment coding, posts were classified into three categories: positive (expressing favorable emotions, optimism, happiness), negative (expressing unfavorable emotions, criticism, sadness), and neutral (expressing neither positive nor negative sentiment). For moral element coding, posts were classified into four categories based on their dominant moral framing according to the Big Three ethics framework: autonomy, community, divinity, and neutral (absence of clear moral content). The gpt-oss-20B and Qwen3-32B models, deployed via Dify with reasoning mode enabled, were employed for both coding tasks.

**Validation of Sentiment and Moral Coding**. To validate the LLM-based coding procedures, we conducted inter-rater reliability analyses comparing human coders with both LLMs across both coding tasks. A randomly selected sample of 200 Weibo posts was independently coded by human coders and both LLM models. One post was excluded due to coding errors, resulting in a final validation sample of 199 posts. Results are shown in Figure 2.

For sentiment analysis, the gpt-oss-20b model achieved substantial agreement with human coders (Cohen's $\kappa = 0.692$, $p < .001$), with 79.9% raw agreement. Category-specific analysis indicated exceptional agreement on positive sentiment (91.0%) and negative sentiment (90.5%), with moderate agreement on neutral sentiment (66.7%). The Qwen3-32b model demonstrated moderate agreement (Cohen's $\kappa = 0.558$, $p < .001$, 71.86% raw agreement), with good but lower performance across all sentiment categories (negative: 76.2%, positive: 74.6%, neutral: 67.8%).

For Big Three moral ethics coding, the gpt-oss-20B model demonstrated substantial agreement with human coders (Cohen's $\kappa = 0.677$, $p < .001$), achieving 83.92% raw agreement. Category-specific analysis revealed excellent agreement on neutral posts (93.8%), strong agreement on community-related posts (75.7%), moderate agreement on autonomy (57.1%), and limited but acceptable agreement on divinity (50.0%, $n = 6$). In contrast, the Qwen3-32b model showed moderate agreement (Cohen's $\kappa = 0.543$, $p < .001$, 76.38% raw agreement), with notably reduced performance on autonomy-related content (39.3% agreement).

Based on these results, the gpt-oss-20b model was selected for automated coding of the full dataset for both sentiment analysis and Big Three moral ethics coding.

**Figure 2**

*Confusion Matrices for Inter-rater Reliability Between Human Coders and AI Models (gpt-oss-20b and Qwen3-32b) on Big Three Morality and Sentiment*

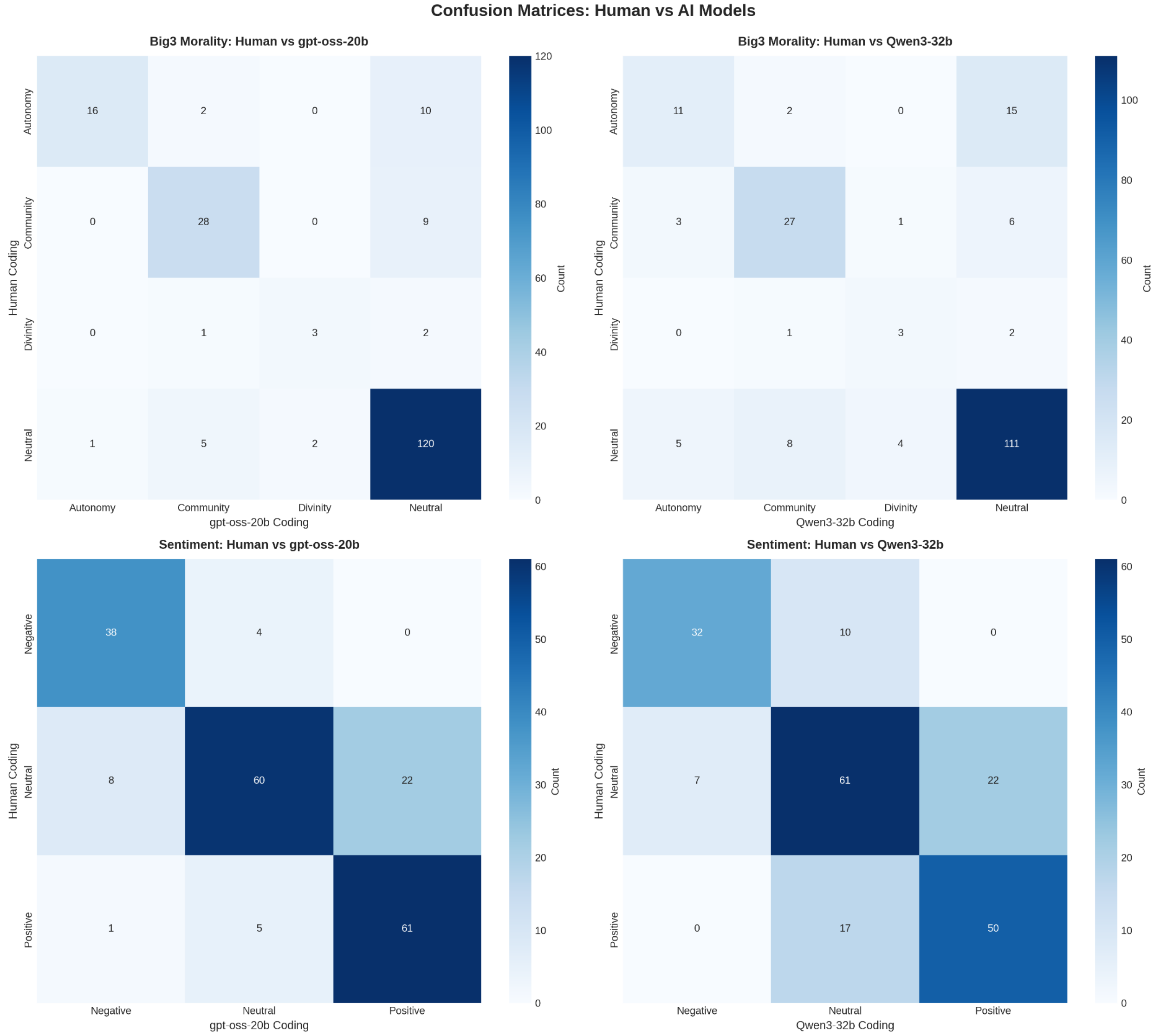

### *Topic modeling*

To explore the semantic structure of marriage-related discourse, we employed word embedding analysis to identify concepts closely associated with marriage across both platforms. Word embeddings capture semantic relationships between words based on their co-occurrence patterns in text, revealing which concepts are closely linked in public conversation and how these associations might differ between platforms.

Text preprocessing was conducted separately for each platform. For Weibo, post content served as the textual input, while Xiaohongshu analysis used post titles given the platform's emphasis on visual content. Preprocessing involved three steps: (1) removing posts with missing content, (2) removing punctuation marks while preserving word boundaries, and (3) performing Chinese word segmentation using the Jieba library (Sun, 2012/2023).

Word embeddings were generated using the FastText algorithm implemented through the Gensim library (Řehůřek, 2022). FastText extends Word2Vec by representing words as character n-grams, enabling robust embeddings for morphologically complex Chinese text (Grave et al., 2018). The FastText model was configured with identical hyperparameters across both datasets to ensure consistency: 500-dimensional vectors to balance semantic richness with computational efficiency, 5-word symmetric context window, skip-gram training with 10 negative samples, and character n-grams of 2-5 characters. Words appearing fewer than 10 times were excluded to filter rare terms. Models were trained for 10 epochs with parallelization across 4 CPU cores.

To identify words semantically related to marriage, we first calculated the centroid vector of the six search terms（婚姻，嫁娶，成亲，成婚，成家，结婚）and identified the top 100 most similar words based on cosine similarity. Hierarchical agglomerative clustering with Ward's linkage was then applied to identify latent semantic groupings within the marriage-related vocabulary. The clustering operated on the full 500-dimensional embeddings without prior dimensionality reduction to preserve semantic information. The number of clusters was determined through dendrogram inspection for natural semantic breaks and evaluation of thematic interpretability. To visualize the high-dimensional semantic relationships, we applied Uniform Manifold Approximation and Projection (UMAP) with 15 nearest neighbors and minimum distance of 0.1, projecting the 500-dimensional word vectors into 2D space for interpretation.

## Results

### RQ1: Sentiment Analysis

Sentiment distributions differed notably between platforms. In the Weibo dataset (see Table 1), positive sentiment was most prevalent (40.29%, n = 28,778), followed by neutral (33.94%, n = 24,246) and negative sentiment (25.77%, n = 18,409). The ratio of positive to negative sentiment was approximately 1.56:1, suggesting that while celebratory and supportive expressions dominated public marriage discourse, critical and distressed expressions remained substantial.

In contrast, Xiaohongshu (see Table 2) showed a predominance of neutral sentiment (59.44%, n = 11,852), with positive sentiment at 30.98% (n = 6,177) and negative sentiment representing only 9.58% (n = 1,910). This pattern suggests that Xiaohongshu users primarily exchange practical information and experiences about marriage rather than expressing strong emotions. When users did express emotions, positive posts outnumbered negative ones by more than three to one (3.24:1).

### RQ2: Moral Elements Analysis

Across both platforms, most marriage-related discourse did not invoke explicit moral elements. In the Weibo dataset (see Table 1), 74.64% of posts (n = 53,317) were morally neutral. Among posts that did employ moral framing, community was most common (13.79%, n = 9,850), followed by autonomy (9.94%, n = 7,100) and divinity (1.64%, n = 1,169). This pattern indicates that when Weibo users explicitly invoked moral reasoning in marriage discussions, they primarily referenced communal values and relational obligations.

The Xiaohongshu dataset exhibited an even stronger pattern of moral neutrality (see Table 2), with 94.22% of posts (n = 18,771) not invoking explicit moral elements. Among the minority of posts with moral framing, community (2.84%, n = 566) and autonomy (2.63%, n = 524) were nearly equally represented, while divinity remained rare (0.31%, n = 61). This distribution indicates that marriage-related discourse on Xiaohongshu focused predominantly on practical and aesthetic considerations rather than moral principles.

**RQ3: Association Between Moral Elements and Sentiment**

A chi-square test of independence revealed a significant relationship between moral elements and sentiment in the Weibo dataset revealed a significant association, $\chi^2(6, n = 71{,}432) = 6954.27$, $p < .001$. The effect size was medium (Cramér's $V = 0.22$), indicating that while the association was present, moral framing explained a small-to-medium portion of sentiment variance. Examination of sentiment distributions across moral categories revealed distinct patterns (See Table 1). Posts invoking autonomy were predominantly negative (50.03%), as were posts invoking community (49.34%). In contrast, posts with divinity-based framing showed the highest proportion of neutral sentiment (50.47%), while morally neutral posts showed the highest proportion of positive sentiment (45.71%). Standardized residuals confirmed these patterns, with the autonomy-negative combination showing over-representation (+40.27), as did the community-negative combination (+46.08). The neutral-positive combination was also over-represented (+19.74), whereas the neutral-negative combination was under-represented (−33.72). Post-hoc pairwise chi-square tests indicated that all Big Three categories differed significantly from one another in their sentiment distributions (all *p*s < .001).

A similar pattern emerged in the Xiaohongshu dataset. Chi-square analysis confirmed a significant association between moral elements and sentiment, $\chi^2(6, n = 19{,}922) = 973.23$, *p*

< .001. Analysis of sentiment distributions revealed patterns largely consistent with Weibo findings (see Table 2). Posts invoking autonomy showed the highest proportion of negative sentiment (38.74%), as did community posts (33.75%). In contrast, divinity posts were mostly positive (47.54%), while morally neutral posts were predominantly neutral (60.65%). Standardized residuals identified strong associations, particularly for the autonomy-negative combination (+21.58) and the community-negative combination (+18.59). The divinity-positive combination showed slightly over-representation (+2.32), while neutral moral framing was associated with neutral sentiment (+2.17). These residuals indicated that posts invoking either individual autonomy or collective obligations were more negative than expected under independence, whereas posts invoking traditional or sacred elements of marriage were more positive than expected.

**Table 1**

*Weibo Frequency Distribution by Big Three Moral Elements and Sentiment*

| Category | Negative | *Row %* | Neutral | *Row %* | Positive | *Row %* | Total |
|---|---|---|---|---|---|---|---|
| Autonomy | **3,552** | *50.03%* | **1,654** | *23.30%* | **1,894** | *26.68%* | **7,100** |
| *Column %* | *19.29%* | | *6.82%* | | *6.58%* | | |
| Community | **4,860** | *49.34%* | **2,846** | *28.89%* | **2,143** | *21.76%* | **9,849** |
| *Column %* | *26.40%* | | *11.74%* | | *7.45%* | | |
| Divinity | **210** | *17.96%* | **590** | *50.47%* | **369** | *31.57%* | **1,169** |
| *Column %* | *1.14%* | | *2.43%* | | *1.28%* | | |
| Neutral | **9,787** | *18.36%* | **19,155** | *35.93%* | **24,372** | *45.71%* | **53,314** |
| *Column %* | *53.16%* | | *79.00%* | | *84.69%* | | |

*Note.* Row % shows percentages within each Big Three category. Column % shows percentages within each sentiment column for autonomy, community, divinity, and neutral categories respectively. Error codes are not included.

**Table 2**

*Xiaohongshu Frequency Distribution by Big Three Moral Elements and Sentiment*

| Category | Negative | *Row %* | Neutral | *Row %* | Positive | *Row %* | Total |
|---|---|---|---|---|---|---|---|
| Autonomy | **203** | *38.7%* | **181** | *34.5%* | **140** | *26.7%* | **524** |
| *Column %* | *10.6%* | | *1.5%* | | *2.3%* | | |
| Community | **191** | *33.7%* | **252** | *44.5%* | **123** | *21.7%* | **566** |
| *Column %* | *10.0%* | | *2.1%* | | *2.0%* | | |
| Divinity | **10** | *16.4%* | **22** | *36.1%* | **29** | *47.5%* | **61** |
| *Column %* | *0.5%* | | *0.2%* | | *0.5%* | | |
| Neutral | **1,503** | *8.0%* | **11,385** | *60.7%* | **5,883** | *31.3%* | **18,771** |
| *Column %* | *78.8%* | | *96.2%* | | *95.3%* | | |

*Note.* Row % shows percentages within each Big Three category. Column % shows percentages within each sentiment column for autonomy, community, divinity, and neutral categories respectively. Error codes are not included.

**RQ4: Topic Modeling**

*Weibo Semantic Clusters*

Hierarchical clustering with Ward's linkage criterion produced a five-cluster solution with considerable size variation (see Figure 3). The largest cluster (n = 42, 39.6%) contained ceremonial and affective terminology including 成亲 (getting married), 喜结良缘 (a happy marriage), and 结婚仪式 (wedding ceremony), representing the romantic and ceremonial dimensions of marriage. The second cluster (n = 35, 33.0%) encompassed core institutional marriage terminology, such as 离了婚 (divorced), 假结婚 (fake marriage), and 家庭婚姻 (family and marriage), suggesting a focus on marriage as a social institution. Notably, this institutional marriage cluster also contained several strongly negative terms, including 万劫不复 (irredeemably lost/beyond redemption), 害人害己 (harming others and oneself), 名存实亡 (existing in name only), indicating that institutional marriage discourse encompasses both family formation and perspectives on marital dysfunction. A third cluster (n = 18, 17.0%) featured Chinese almanac terminology related to auspicious activities and traditional fortune-telling practices, including 纳婿 (taking a son-in-law), 纳彩 (betrothal gifts), and 求嗣忌 (seeking offspring taboo). The fourth cluster includes several words related to astrology, and the final cluster was considered insignificant due to its very small number of words.

**Figure 3**

*Hierarchical clustering showing a 5-cluster solution for Weibo Embedding*

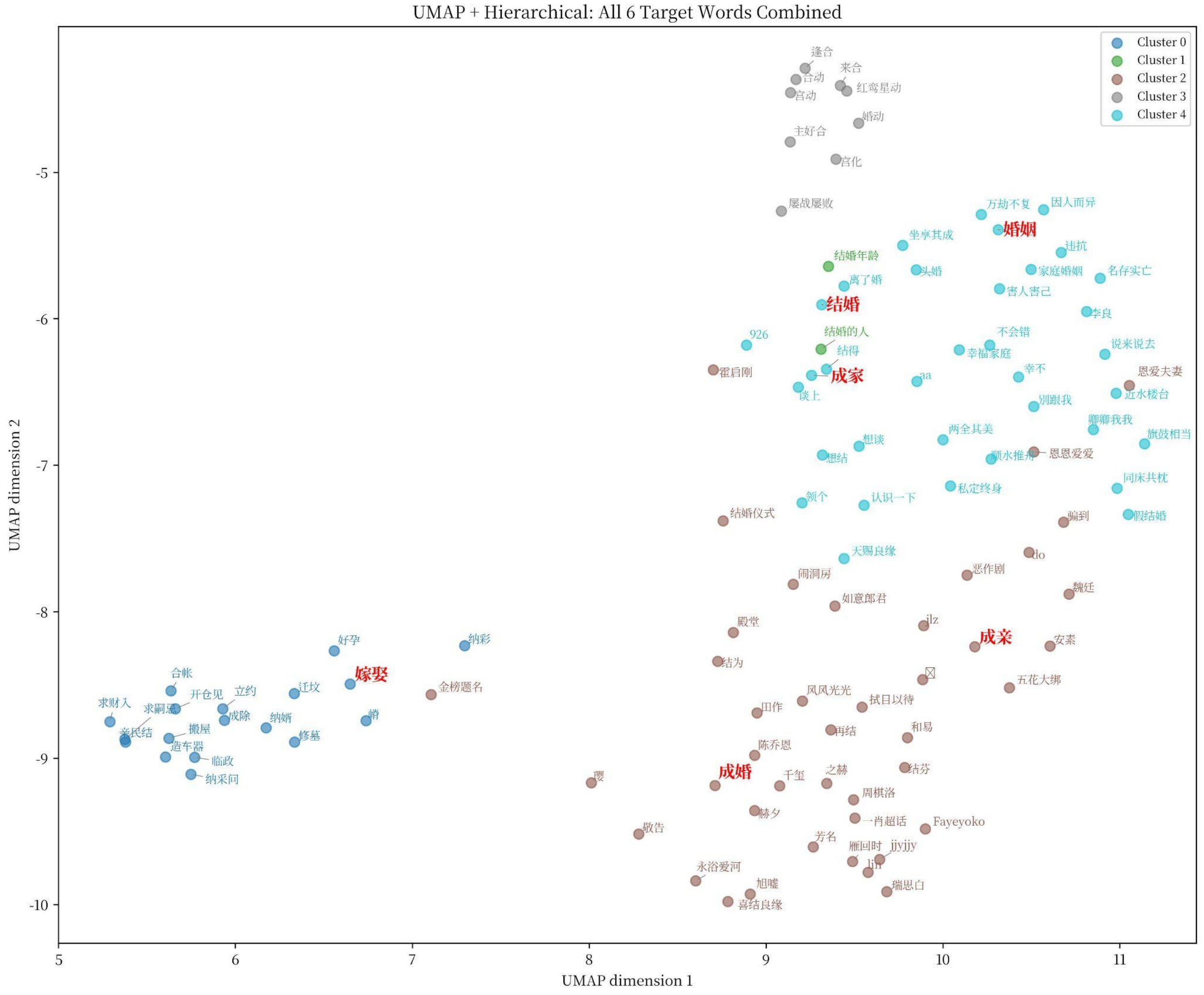

Xiaohongshu embeddings showed notably different semantic organization compared to Weibo. Hierarchical clustering identified five clusters with markedly unbalanced distributions (see Figure 4). The largest cluster (n = 50, 47.2%) represented relationship processes and experiential dimensions, including relationship stages such as 恋爱 (dating), 分手 (breaking up), 闪婚 (flash marriage), emotional states such as 愿意 (willing), 真心 (sincere). The second cluster (n = 47, 44.3%) diverse marriage-related concepts including emotional states 遗憾 (regret), 清醒 (clear-headed), 不敢 (dare not); material considerations such as 房子 (house), 城

市 (city); identity markers such as 身份 (identity), 男方 (the groom's side). An intermediate cluster (n = 6, 5.7%) contained formal ceremony terms that appeared to co-occur with narrative elements, suggesting usage in fictional or dramatic contexts. The remaining two clusters were very small (n = 2, 1.9% and n = 1, 0.9%), containing brief descriptions or "miscellaneous terms".

**Figure 4**

*Hierarchical clustering showing a 5-cluster solution for Xiaohongshu Embedding*

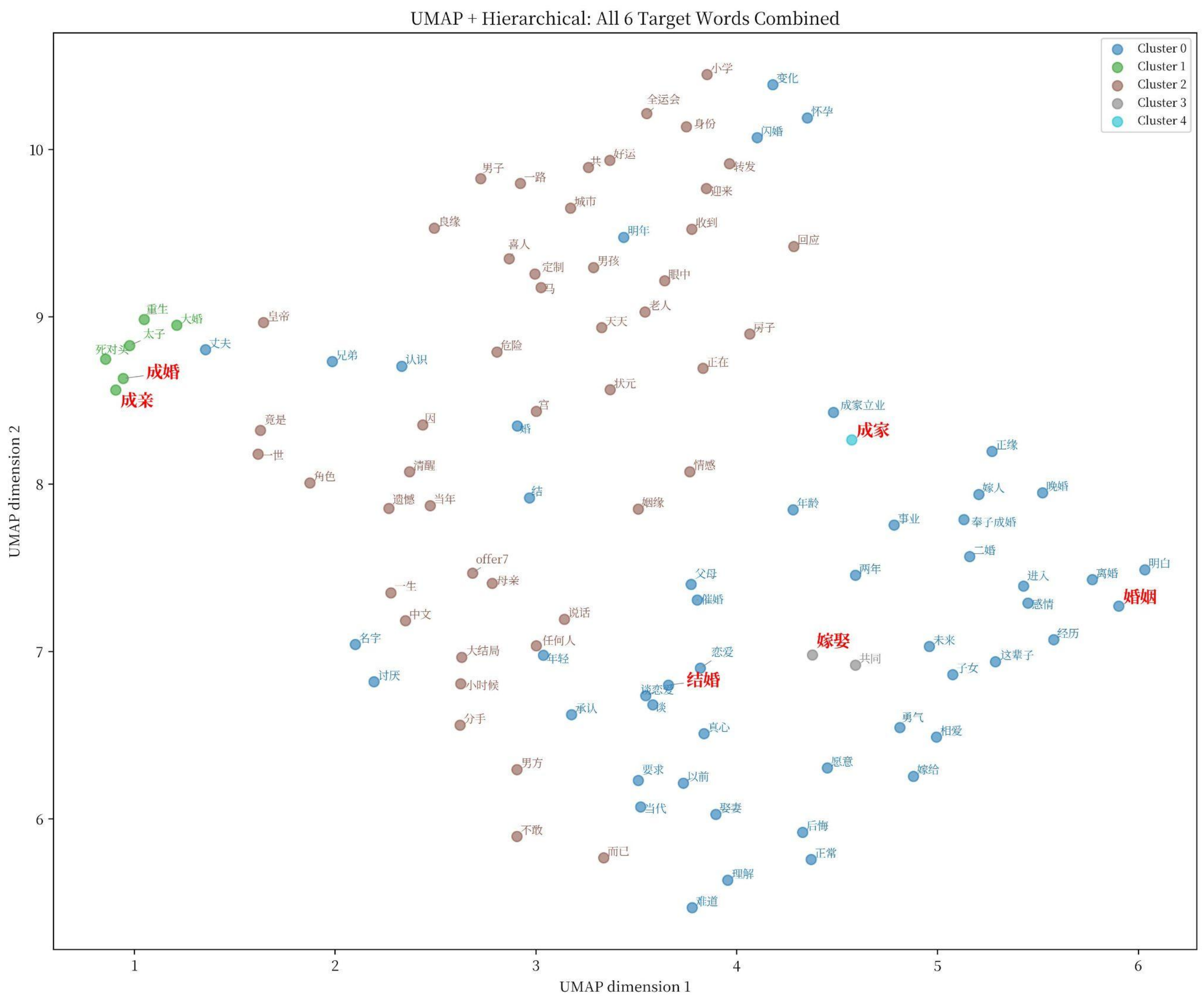

The semantic structures differed substantially between platforms. Weibo discourse exhibited balanced organization across ceremonial (39.6%), institutional (33.0%), and traditional (17.0%) domains. In contrast, Xiaohongshu discourse was dominated by experiential and relational content (91.5% combined in the two largest clusters), with formal ceremonial language appearing primarily in narrative contexts rather than forming an independent institutional domain.

## Discussion

By examining how moral elements related to sentiment in naturalistic discourse, we extended previous survey-based research and demonstrated the moral elements underlying public evaluations of marriage in China. Our findings showed how different moral elements related to marriage attitudes, offering insights for developing culturally informed policies and communication strategies that address the concerns driving marriage decline in Chinese contexts.

### The Negativity in Chinese Marriage Discourse

This study revealed a pattern across two major Chinese social media platforms: when users invoked moral elements to discuss marriage, they expressed negative sentiment. This association was stronger for posts employing autonomy-based and community-based moral elements. On Weibo, posts invoking autonomy were mostly negative (50.03%), as were community-based posts (49.34%). This pattern appeared on Xiaohongshu as well, where autonomy posts showed 38.74% negative sentiment and community posts 33.75% negativity. In contrast, morally neutral posts were more positive on both platforms.

The consistency of this pattern across platforms, despite differences in user demographics and platform features, suggested that moral discourse about marriage in contemporary China primarily expressed critique and distress rather than celebration or affirmation. The negativity of autonomy-based moral discourse reflected perceptions that marriage violated individual rights, constrained personal freedom, and prioritized external obligations over individual wellbeing. Posts employing autonomy likely invoked concepts of harm, coercion, unfairness, and the suppression of personal choice. When individuals perceived violations of personal autonomy, they responded with anger and moral outrage (Rozin et al., 1999). This pattern was particularly notable in the Chinese cultural context, where autonomy-based moral reasoning had historically been less emphasized than community and divinity ethics emphasizing familial obligation and social harmony. The prevalence of autonomy-based critique signaled a shift in moral thinking among younger, urban, educated populations who constituted the main users of platforms like Weibo and Xiaohongshu. These users evaluated marriage not through traditional values of duty and propriety, but through perspectives that emphasized individual rights, consent, and personal fulfillment. When marriage failed to meet these standards, users expressed moral outrage about autonomy violations.

The topic modelling analysis reinforces this interpretation. On Weibo, the institutional marriage cluster contained not only terms like “family and marriage” and “happy family”, but also strongly negative terms including “irredeemably lost/beyond redemption”, “harming others and oneself”, “existing in name only”, and “deceived/tricked”. This combination of positive family language with language of deception and harm suggested that institutional marriage discourse included experiences of betrayal and distress. Similarly, on Xiaohongshu, the dominant experiential cluster included relationship processes alongside terms like “regret”, “breakup”, and “pressure to marry”, indicating that personal narratives of marriage were associated with conflict, coercion, and loss.

This pattern had important implications. When moral discourse about marriage became negative and emphasized autonomy violations, marriage shifted from a valued social institution to a perceived threat to individual wellbeing. This transformation had consequences: China's marriage rate declined substantially, dropping 54.67% from 2013 to 2024 (National Bureau of Statistics, 2025). The moral discourse documented in this study may have both reflected and reinforced these behavioral trends, creating a cycle in which negative views of marriage deterred participation, which in turn reinforced the perception that those who did marry had been coerced or made poor choices.

**Cross-Platform Differences**

Despite the consistent association between moral framing and negativity across platforms, differences in topic patterns suggested that marriage discourse was becoming fragmented across demographic and platform lines. Weibo's marriage discourse included diverse moral framings (25.37% of posts invoked moral elements) and showed distinctions between institutional, ceremonial, and traditional topics. The presence of clusters for ceremonial affect, traditional almanac practices, and astrological prediction indicated that Weibo users engaged with marriage in multiple ways, from romantic celebration to traditional ritual to institutional critique. In contrast, Xiaohongshu's marriage discourse was predominantly neutral and characterized by two experiential clusters that included relationship processes, emotional states, and practical concerns. The platform's emphasis on visual content and consumer culture appeared to favor practical and personal representations of marriage as individual choice and lifestyle decision. These platform differences likely reflected demographic differences and different platform features (QuestMobile, 2024). Weibo's diverse user base and public discussion format may have encouraged more varied moral discourse and debate, while Xiaohongshu's younger, predominantly female user base and focus on lifestyle content may have favored practical, experience-sharing content over moral debate.

**Implications**

This study aimed to understand not only whether people view marriage positively or negatively, but also the moral reasoning that underlies these attitudes. This distinction matters for policy and intervention design. China's declining marriage and birth rates have prompted government initiatives aimed at "promoting marriage culture" and combating "negative attitudes toward marriage and childbearing." However, our findings revealed why such interventions may face difficulties.The analysis showed that negative attitudes toward marriage stemmed from two distinct moral concerns, each requiring a different approach. First, posts invoking autonomy concerns about violations of individual rights, personal freedom, and unfair treatment in marriage. These users viewed marriage as constraining their choices and sacrificing their wellbeing to family obligations. For this group, interventions should address the structural conditions that make marriage feel coercive, such as discriminatory workplace policies, unequal domestic labor divisions, and social pressures that limit individual agency. Simply promoting "marriage culture" or appealing to family duty would likely provoke further resistance, as these messages would be perceived as attempts at coercion. Second, posts invoking community elements expressed concerns about oppressive family obligations and burdensome social expectations. These users viewed marriage as imposing unrealistic demands from extended family, such as eldercare responsibilities, financial support obligations, or rigid gender role expectations. For this group, interventions should address how family and society create pressures on married couples, such as inadequate public support for eldercare, lack of flexible work arrangements, or inflexible traditional role expectations. Third, our findings suggested that the most positive marriage discourse was morally neutral, practical, and focused on lifestyle considerations. This suggests that some individuals may be more responsive to approaches that emphasize practical benefits, personal compatibility, and aesthetic/lifestyle aspects of partnership rather than moral arguments about duty or fulfillment.

These findings indicate that general pro-marriage campaigns risk ineffectiveness without addressing underlying moral concerns. Effective interventions must be targeted, addressing the specific structural barriers different groups experience rather than relying on broad cultural appeals.

**Limitations and Future Directions**

This study's reliance on social media data, though advantageous, also has some limitations. First, Weibo and Xiaohongshu users were predominantly young, urban, educated, and female, a demographic more likely to delay or avoid marriage. The documented negativity may have reflected sampling limitations rather than population-wide attitudes. However, this limitation is also a strength: these platforms capture the moral discourse of the demographic group whose marriage behavior will shape China's demographic future over the coming decades. Their attitudes, even if not representative, may predict future trends. Second, the cross-sectional design could not establish causal relationships between moral framing and sentiment. Do negative experiences lead to moral framing, or does moral framing increase negative feelings? Longitudinal studies tracking individuals' discourse over time, or experiments testing moral framing in marriage scenarios, could clarify these causal relationships. Finally, the study's focus on public social media posts may have missed private conversations, family discussions, and offline communication that might have revealed different patterns. Future research using interviews, observational studies, and analysis of private messages could provide a fuller understanding of how Chinese people morally reason about marriage. Additionally, comparative research examining similar patterns in other developing societies with declining marriage rates could clarify whether the patterns documented here represent a common feature of demographic change or a culturally specific Chinese pattern.

## Conclusion

This study revealed that moral discourse about marriage on Chinese social media was predominantly negative when users invoked autonomy and community elements. Divinity-framed posts showed positive sentiment but were rare, while the most positive discourse was morally neutral. Given this lack of moral consensus, increasing marriage rates through messaging campaigns alone appears unlikely. The study suggests that the challenge facing Chinese society was not simply to promote marriage but to establish moral consensus about family life in an era of diverse values and individual choice.